\documentclass[12pt] {article}
\usepackage{epsfig}
\usepackage{amstex}
\newcounter{nombre}
\def\agt{
\mathrel{\raise.3ex\hbox{$>$}\mkern-14mu\lower0.6ex\hbox{$\sim$}}
}
\def\alt{
\mathrel{\raise.3ex\hbox{$<$}\mkern-14mu\lower0.6ex\hbox{$\sim$}}
}

\begin {document}

\bibliographystyle{unsrt}    

\Large
{\centerline{\bf Expectation values of local fields}}
\vspace{3mm} 
{\centerline{\bf for a two-parameter family of integrable}
\vspace{3mm} 
{\centerline{\bf models and related perturbed}
\vspace{3mm}
{\centerline{\bf conformal field theories}
\vspace{3mm} 

\large

\vspace{7mm}

\centerline {\bf P.~Baseilhac, V. A. Fateev\,\footnote{On leave of absence from L. D. Landau Institute for Theoretical Physics, ul. Kosygina 2, 117940 Moscow, Russia}}

\small\normalsize

\vspace{3mm}
\centerline {Laboratoire de Physique Math\'ematique, Universit\'e Montpellier II}
\centerline {Place E.~Bataillon, 34095 Montpellier, France}
\vspace{3mm}

\vspace{6mm}
\large
\begin{abstract}
We calculate the vacuum expectation values of local fields for the two-parameter family of integrable field theories introduced and studied in \cite{11}. Using this result we propose an explicit expression for the vacuum expectation values of local operators in parafermionic sine-Gordon models and in integrable perturbed $SU(2)$ coset conformal field theories.
\end{abstract}
\normalsize
\newpage

\section{Introduction}
\label{introduction} 

\ \ \ \ The vacuum expectation values (VEV) of local fields play an important role in quantum field theory (QFT) and statistical mechanics \cite{1,2}. In statistical mechanics the VEVs determine the ``generalized susceptibilities'' i.e. the linear response of the system to external fields. In the QFT defined as a perturbed conformal field theory (CFT) the VEVs provide all the information about its correlation functions that is not accessible through direct calculations in conformal perturbation theory \cite{3}. Recently some progress was made in the calculations of the VEVs in two dimensional integrable QFT. In the ref. \cite{4}, an explicit expression for the VEVs of the exponential fields in the sine-Gordon and sinh-Gordon models was proposed. It was shown in \cite{5} that this expression can be obtained as the minimal solution of certain ``reflection relations'' which involve the Liouville ``reflection amplitude'' \cite{6}. In this approach the sinh-Gordon QFT can be considered as a perturbed Liouville CFT. In the same paper the ``reflection amplitude'' for the boundary Liouville theory was used for the calculation of the VEVs for the boundary sinh-Gordon and sine-Gordon theories. In ref. \cite{7} the same method was applied to the calculation of the VEVs of the exponential fields in the so called Bullough-Dodd model with real and purely imaginary coupling. It is known that $c<1$ minimal CFT perturbed by the operators $\Phi_{12}$ and $\Phi_{13}$ can be obtained by a quantum group (QG) restriction of the sine-Gordon  \cite{8} and imaginary  Bullough-Dodd \cite{9} models with special values of the coupling. These relations were used in \cite{7} to calculate the VEVs of primary fields in a perturbed minimal CFT. The results are in good agreement with numerical data obtained in \cite{10}.

In this paper we consider the VEVs of local fields in a two-parameter family of integrable QFTs with the action :
\begin{eqnarray}
{\cal{A}} &=& \int d^2x \Big[ \frac{1}{16\pi} \big[ (\partial_\mu\varphi_1)^2 + (\partial_\mu\varphi_2)^2 + (\partial_\mu\varphi_3)^2\big] \nonumber \\
&-& 2\mu  \big[ e^{\alpha \varphi_1}\cos(\beta\varphi_2+\gamma\varphi_3) + e^{-\alpha \varphi_1}\cos(\beta\varphi_2-\gamma\varphi_3)\big]\Big],
\label{a1}
\end{eqnarray}
where the parameters $\alpha, \beta, \gamma$ satisfy the relation :
\begin{eqnarray}
 \beta^2+\gamma^2-\alpha^2=\frac{1}{2}.
\label{cond}
\end{eqnarray}
In this paper we use the following parametrization for $\alpha, \beta, \gamma$ :
\begin{eqnarray}
4\alpha^2=p+n,\ \ \ 4\beta^2= n+2,\ \ \  4\gamma^2=p.
\end{eqnarray}

The QFT (\ref{a1}) was introduced and studied in \cite{11} where the ``dual'' $\sigma$-model representation for this theory available in the strong coupling region $\alpha, \beta, \gamma >> 1$ was constructed.\\
The QFT (\ref{a1}) contains as particular cases the known integrable models : $N=2$ supersymmetric sine-Gordon theory, $O(4)$ and $O(3)$ non-linear $\sigma$-models, the sausage model \cite{12}, \cite{13} and others. In ref. \cite{14} the QFT (\ref{a1}) was used for the study of tunneling in quantum wires. For integer $n$ the theory (\ref{a1}) admits a QG restriction to the parafermionic sine-Gordon models \cite{15} with actions (\ref{a2}). For integer $n$ and $p$ it can be restricted to the integrable perturbed $SU_2(n)\otimes SU_2(p-2)/SU_2(p+n-2)$ coset CFT \cite{16}.

In section 2 we use CFT methods to calculate the ``reflection amplitudes'' for the exponential fields $\exp(a\varphi_1+ib\varphi_2+ic\varphi_3)$ in the QFT (\ref{a1}). We solve the functional ``reflection relations'' for the VEVs of these fields and do some tests to confirm the result. The strong coupling asymptotics of the VEVs are considered. In section 3 we use our results for the QFT (\ref{a1}) to calculate the VEVs of the basic fields in the parafermionic sine-Gordon models. As particular cases we consider $N=1$ and restricted $N=2$ supersymmetric sine-Gordon models. In section 4 we calculate the VEVs of the primary fields in the integrable perturbed $SU(2)$ coset CFT models. As an example, we consider a minimal $N=1$  supersymmetric CFT perturbed by an operator preserving supersymmetry. In the last section we discuss the relation between VEVs and certain quantities in CFT.
  
\section{Reflection amplitudes and vacuum expectation values in two parameter family of integrable QFT}

\ \ \ \ \ The starting point for the calculation of the VEVs in an integrable QFT, which can be defined as a perturbed CFT, is the calculation of the ``reflection amplitudes'' for the local fields in the basic CFT \cite{5,7}. For the QFT (\ref{a1}) we can interpret the first four terms in eq. (\ref{a1}) as the action of the CFT and the last one as the perturbation (the choice of perturbation between two last terms is, of course, conventional). The action of the corresponding CFT has the form :
\begin{eqnarray}
{\cal{A}}_{CFT} = \int d^2x \Big[ \frac{1}{16\pi} \big[ (\partial_\mu\varphi_1)^2 + (\partial_\mu\varphi_2)^2 + (\partial_\mu\varphi_3)^2\big] -2\mu e^{\alpha \varphi_1}\cos(\beta\varphi_2+\gamma\varphi_3)\Big],
\label{aCFT}
\end{eqnarray}
where parameters $\alpha, \beta, \gamma$ satisfy the condition (\ref{cond}). This action can be also obtained as the ``conformal'' limit ($\mu \rightarrow 0$, $\varphi_1 \rightarrow \infty$, $\mu e^{\alpha \varphi_1}$ is fixed.) of the action (\ref{a1}).

The holomorphic stress-energy tensor $T(z)$, where $z=x_1+ix_2$, $\overline{z}=x_1-ix_2$ are the complex coordinates of $\mathbb{R}^2$ : 
\begin{eqnarray}
T(z)=-\frac{1}{4}(\partial_z\varphi_1)^2 -\frac{1}{4}(\partial_z\varphi_2)^2 -\frac{1}{4}(\partial_z\varphi_3)^2 + q_{\alpha}\partial^2_z\varphi_1,\ \ \mbox{with} \ \ q_{\alpha}=\frac{1}{4\alpha}
\end{eqnarray}
ensures the local conformal invariance of the QFT  (\ref{aCFT}) with the central charge $c=3+\frac{3}{2\alpha^2}$. The exponential fields 
\begin{eqnarray}
V(a,b,c;x) = \exp(a\varphi_1+ib\varphi_2+ic\varphi_3) \ , \ \ \ V^{\dagger}(a,b,c;x)=V(a,-b,-c;x)
\end{eqnarray}
are spinless conformal primary fields with dimensions :
\begin{eqnarray}
\Delta=\Delta(a,b,c)=a(2q_{\alpha}-a)+b^2+c^2.
\label{weight}
\end{eqnarray}
In particular, fields 
\begin{eqnarray}
V(\alpha,\beta,\gamma;x) = \exp(\alpha\varphi_1+i\beta\varphi_2+i\gamma\varphi_3)\ , \ \ \ V^{\dagger}(\alpha,\beta,\gamma;x)=V(\alpha,-\beta,-\gamma;x)
\end{eqnarray}
have conformal dimension equal to one.\\
Besides the conformal symmetry generated by $T\equiv T_2$, the QFT  (\ref{aCFT}) possesses an additional infinite-dimensional symmetry generated by the chiral algebra ${\mathfrak{T}}$ which includes an infinite number of holomorphic fields $T_2, T_3,...$ with integer spins. The detailed description of the chiral algebra ${\mathfrak{T}}$ is not within the scope of this paper. As an example of the spin 3 field $T_3\in{\mathfrak{T}}$, we give here the holomorphic primary field : 
\begin{eqnarray}
T_3 = D(\partial_z\phi)^3\ +\ E(\partial_z\phi)(\partial_z\varphi_1)^2\ + \ F(\partial_z\phi)(\partial_z^2\varphi_1)\ + \ G(\partial_z^2\phi)(\partial_z\varphi_1) + \ L(\partial_z^3\phi),
\end{eqnarray}
where $\phi = \beta\varphi_2 + \gamma\varphi_3$, \ $D=\frac{1+6\alpha^2}{1+2\alpha^2}$;\  $E=3\alpha^2$;\  $F=6\alpha^3$;\  $G= -3\alpha(1+2\alpha^2)$;\  $L=\frac{1}{2}(1+2\alpha^2)$.\\
The other fields $T_j$, $j>3$ can be obtained by fusion of the field $T_3$.

The primary fields $\Phi_t$ of the chiral algebra ${\mathfrak{T}}$ can be classified by the eigenvalues $t_j$ of the operators $T_{j,0}$ (the zero Fourier components of the currents $T_j$) :
\begin{eqnarray}
T_{j,0}\Phi_t=t_j\Phi_t, \ \ \ \ \ t_2=\Delta.
\end{eqnarray}
The fields $V(a,b,c)$ are the primary fields of the chiral algebra ${\mathfrak{T}}$ with the parameters $t_j=t_j(a,b,c)$. All of the functions $t_j(a,b,c)$ possess the reflection symmetry $t_j(a,b,c)=t_j(2q_{\alpha}-a,b,c)$. For $j=2$, this follows from eq. (\ref{weight}); for $j=3$ one can easily derive this property from the explicit form of $T_3$. The fields $V(a,b,c)$ and $V(2q_{\alpha}-a,b,c)$ are the reflection image of each other and are related by the linear transformation : 
\begin{eqnarray}
V(a,b,c;x) = R_{\alpha}(a,b,c)V(2q_{\alpha}-a,b,c;x),
\end{eqnarray}
where $R_{\alpha}(a,b,c)$ is the ``reflection amplitude''.

This function is an important object in CFT (it defines the two-point functions of the operators $V$) and plays a crucial role in the calculation of one-point functions in perturbed CFT. To calculate the function $R_{\alpha}(a,b,c)$ we introduce the fields $\Phi_t$ :
\begin{eqnarray}
\Phi_t=N^{-1}(a,b,c) V(a,b,c),
\label{field}
\end{eqnarray}
where the normalization constant $N(a,b,c)=N(a,-b,-c)$ is chosen in the way that fields $\Phi_t$ satisfy the conformal normalization condition\,\footnote{This condition will be used later for the normalization of operators in perturbed CFT.} :
\begin{eqnarray}
<\Phi^{\dagger}_t(x)\Phi_t(y)> = \frac{1}{|x-y|^{4\Delta}}.
\label{correl}
\end{eqnarray}
The normalized fields $\Phi_t$ are invariant under the reflection transformation and hence : 
\begin{eqnarray}
R_{\alpha}(a,b,c)=\frac{N(a,b,c)}{N(2q_{\alpha}-a,b,c)}.
\label{reflec}
\end{eqnarray}
For the calculation of $R_{\alpha}(a,b,c)$, we note that operators $Q_{+},Q_{-}$ defined as : 
\begin{eqnarray}
Q_{+}=\mu \int d^2x V(\alpha,\beta,\gamma;x)\ \ \ ; \ \ \ Q_{-}=\mu \int d^2x V(\alpha,-\beta,-\gamma;x)
\end{eqnarray}
commute with all of the elements of the chiral algebra ${\mathfrak{T}}$ and can be used as screening operators for the calculation of correlation functions in the CFT (\ref{aCFT}).

For $a=-m\alpha$ with integer $m$, we obtain from eqs. (\ref{field}),(\ref{correl}) the following expression for the function $N(a,b,c)$ in terms of Coulomb integrals : 
\begin{eqnarray}
N^2(a,b,c) = \frac{|x|^{4\Delta}}{(m!)^2}<V(a,-b,-c;0) V(a,b,c;x)Q_{+}^m Q_{-}^m>,
\label{norm}
\end{eqnarray}
where the expectation value in (\ref{norm}) is taken over the Fock vacuum of massless fields $\varphi_1,\varphi_2,\varphi_3$, with correlation functions :
\begin{eqnarray}
<\varphi_i(x) \varphi_j(y)> = -2\delta_{ij} \log |x-y|^2.
\end{eqnarray}
Exactly the same expression (\ref{norm}) for $N^2$ can be obtained if we do perturbation theory in $\mu$ with the action (\ref{aCFT}).

For the calculation of the integral (\ref{norm}) we can use the ``duality'' relation between a $2n$-fold integral with the measure :
\begin{eqnarray}
d\mu_n(u) = \frac{\pi^{-n}}{n!}\prod_{1\leq i <j}^{n}(u_i - u_j)^2 \prod_{j=1}^{n}d^2u_i
\end{eqnarray}
and a $2l$-fold integral with the measure :
\begin{eqnarray}
d\mu_l(v) = \frac{\pi^{-l}}{l!}\prod_{1\leq i <j}^{l}(v_i - v_j)^2 \prod_{j=1}^{l}d^2v_i,
\end{eqnarray}
which has the form :
\begin{eqnarray}
\int d\mu_n(u) \prod_{i=1}^{n}\prod_{j=1}^{l+n+1}|u_i-y_j|^{2a_j-1} = \int d\mu_l(v) \prod_{i=1}^{l}\prod_{j=1}^{l+n+1}|v_i-y_j|^{-2a_j-1} \\
 \times \gamma(\frac{l-n+1}{2}-\sum_{j=1}^{n+l+1}a_j)\prod_{i=1}^{l+n+1}\gamma(\frac{1}{2}+a_i)\prod_{i<j}|y_i-y_j|^{2a_i+2a_j} \nonumber,
\end{eqnarray}
where
\begin{eqnarray}
\gamma(x)=\frac{\Gamma(x)}{\Gamma(1-x)}.
\label{gamma}
\end{eqnarray}
Using this relation we can reduce the integral (\ref{norm}) to a standard integral calculated in \cite{17}. As the result we obtain that for $a=-m\alpha$, the function $R_{\alpha}$ defined by eq. (\ref{reflec}) has the form :
\begin{eqnarray}
R_{\alpha}(a,b,c)&=&\Big(\frac{\pi \mu}{4\alpha^2}\Big)^{-2\overline{A}/\alpha^2}\frac{\Gamma(\frac{1}{2}-2\overline{A}+2B+2C)}{\Gamma(\frac{1}{2}+2\overline{A}+2B+2C)} \frac{\Gamma(\frac{1}{2}-2\overline{A}-2B-2C)}{\Gamma(\frac{1}{2}+2\overline{A}-2B-2C)} \nonumber \\
&& \ \ \ \ \ \ \ \ \ \ \ \ \ \ \ \ \ \ \ \ \ \times \ \ \frac{\Gamma(1+4\overline{A})}{\Gamma(1-4\overline{A})}\frac{\Gamma(1+\overline{A}/\alpha^2)}{\Gamma(1-\overline{A}/\alpha^2)} .
\label{reflec0}
\end{eqnarray}
where
\begin{eqnarray}
A=\alpha a,\ \ \  B=\beta b,\ \ \   C=\gamma c \ \  ; \ \ \  \overline{A}=\alpha(a-q_{\alpha}).
\end{eqnarray}
We accept eq. (\ref{reflec0}) as the proper analytical continuation of the function $R_{\alpha}$ for all $a$.

We define the function $G(a,b,c)$ as the VEV of the operator $V(a,b,c;x)$ in the QFT (\ref{a1}) :
\begin{eqnarray}
G(a,b,c) = <\exp(a\varphi_1+ib\varphi_2+ic\varphi_3)>.
\label{VEV1}
\end{eqnarray}
As was shown in papers \cite{5,7} the VEV (\ref{VEV1}) satisfies the same ``reflection relations'' as the operator $V(a,b,c)$ in the QFT (\ref{aCFT}). In this way, we arrive at a functional equation for the function $
G(a,b,c)$ :
\begin{eqnarray}
G(a,b,c) = R_{\alpha}(a,b,c)G(2q_{\alpha}-a,b,c).
\end{eqnarray}
To obtain functional equations with respect to the variables $b,c$, we note that action (\ref{a1}) is invariant under the transformation $\alpha \leftrightarrow i\beta \ (\alpha \leftrightarrow i\gamma)$. The function $
G(a,b,c)$ is invariant under this transformation together with the substitution $a \leftrightarrow ib \ (a \leftrightarrow ic)$. As a result, the function $G$ satisfies also the functional relation :
\begin{eqnarray}
G(a,b,c) = R_{\beta}(a,b,c)G(a,2q_{\beta}-b,c),
\end{eqnarray}
with
\begin{eqnarray}
R_{\beta}(a,b,c)&=& \Big(\frac{\pi \mu}{4\beta^2}\Big)^{-2\overline{B}/\beta^2}\frac{\Gamma(\frac{1}{2}+2\overline{B}-2A+2C)}{\Gamma(\frac{1}{2}-2\overline{B}-2A+2C)} \frac{\Gamma(\frac{1}{2}+2\overline{B}+2A-2C)}{\Gamma(\frac{1}{2}-2\overline{B}+2A-2C)} \nonumber \\
&& \ \ \ \ \ \ \ \ \ \ \ \ \ \ \ \ \ \ \ \ \ \times \ \ \frac{\Gamma(1-4\overline{B})}{\Gamma(1+4\overline{B})}\frac{\Gamma(1+\overline{B}/\beta^2)}{\Gamma(1-\overline{B}/\beta^2)},
\label{reflec1}
\end{eqnarray}
where 
\begin{eqnarray}
\overline{B}=\beta(b-q_{\beta}); \ \ q_{\beta} = -\frac{1}{4\beta},\nonumber
\end{eqnarray}
and a similar relation with respect to reflection of the parameter $c$.

The minimal solution of these three functional equations has the form : 
\begin{eqnarray}
G(a,b,c) =(\pi \mu)^{2b^2+2c^2-2a^2} \Big[ \frac{\gamma(\frac{1}{2}+P-2A)\gamma(\frac{1}{2}+P-2B)\gamma(\frac{1}{2}+P-2C)}{\gamma(\frac{1}{2}+P)}\Big]^{\frac{1}{2}} \exp{Q},
\label{VEV2}
\end{eqnarray}
where $P=A+B+C$ and 
\begin{eqnarray}
\label{29} 
Q(a,b,c) &=& \int \frac{dt}{t} \Big[ \frac{1}{\sinh 2t} \Big( \frac{\sinh^24Bt}{\tanh 4\beta^2t} + \frac{\sinh^24Ct}{\tanh 4\gamma^2t} - \frac{\sinh^24At}{\tanh 4\alpha^2t}\Big)  \\ 
&& \ \ \ \ \ \ \ \ \ \ \ \ \ \ \ \ \ \ \ \ \ \ \ - \ \ 2(b^2+c^2-a^2)e^{-2t}\Big]\  + \ \ Q_1(a,b,c),\nonumber
\end{eqnarray}
here 
\begin{eqnarray}
Q_1(a,b,c) = \int \frac{dt}{t} \frac{2 \sinh 2Pt \sinh 2(P-2A)t \sinh 2(P-2B)t \sinh 2(P-2C)t}{\sinh^2 t}.\label{Q_1}
\end{eqnarray}
The VEV here is expressed in terms of the parameter of the action $\mu$. To express it in terms of the masses of physical particles we note that QFT (\ref{a1}) possesses a $U(1) \otimes U(1)$ symmetry. Four basic particles of the theory $A_{q_2}^{q_1}(\theta)$ ($\theta$ denotes the rapidity of the particles) have $U(1) \otimes U(1)$ charges $q_1,q_2 = \pm 1$. The full spectrum and the scattering theory of QFT (\ref{a1}) are described in ref. \cite{11}. In particular, it was shown there that the exact relation between the parameter $\mu$ and the mass $M$ of a particle $A_{q_2}^{q_1}$ has the form :
\begin{eqnarray}
\pi \mu = \frac{\pi M}{2} \frac{\Gamma(2\beta^2+2\gamma^2)}{\Gamma(2\beta^2)\Gamma(2\gamma^2)} = \frac{\pi M}{2} \frac{\Gamma(\frac{p+n+2}{2})}{\Gamma(\frac{p}{2})\Gamma(\frac{n+2}{2})}.
\end{eqnarray}

Using eq. (\ref{VEV2}) we can calculate the vacuum bulk energy in the QFT (\ref{a1}) and compare it with the value obtained in \cite{11} by the Bethe Ansatz method. The specific bulk vacuum energy ${\cal{E}}_0(\mu,\alpha,\beta,\gamma)$ is related to the VEV as :
\begin{eqnarray}
-\frac{\partial{\cal{E}}_0}{\partial \mu} = -<\frac{\partial{\cal{L}}}{\partial \mu}> = 4 
G(\alpha,\beta,\gamma),
\label{benergy0}
\end{eqnarray}
where ${\cal{L}}$ is the Lagrangean density of our theory and in the last equation (\ref{benergy0}) we took into account the symmetries of the action (\ref{a1}). The integrals (\ref{29},\ref{Q_1}) for the $G(\alpha,\beta,\gamma)$ can be evaluated explicitly and we obtain :
\begin{eqnarray}
-\frac{\partial{\cal{E}}_0}{\partial \mu} = 2\pi \mu \frac{\Gamma(-\alpha^2)\Gamma(\beta^2)\Gamma(\gamma^2)}{\Gamma(1+\alpha^2)\Gamma(1-\beta^2)\Gamma(1-\gamma^2)},
\label{benergy}
\end{eqnarray}
in exact agreement with the result of ref. \cite{11}.

One can do some other checks of eq. (\ref{VEV2}). A simple consistency check is based on the fact that at the point $\alpha^2=0$, $\beta^2=\gamma^2=\frac{1}{4}$ the QFT (\ref{a1}) reduces to two sine-Gordon models at the free-fermion point. It is easy to see that the functions $G(0,a,b)$ coincide with the expectation values of the exponential operators for the free-fermion theory. One can do a perturbative expansion near this point and find an exact agreement with eq. (\ref{VEV2}). Another check can be done at the point $\alpha=\frac{i}{2}, \beta = \gamma =\frac{1}{\sqrt{8}}$. At this point, the QFT (\ref{a1}) coincide with an imaginary $A_3$ affine Toda field theory with a coupling constant corresponding to a perturbing-operator dimension equal to $\frac{1}{2}$. The comparison of eq. (\ref{VEV2}) with the VEV of $A_3$ Toda theory gives an exact coincidence. At $\gamma^2=\frac{1}{2}$ $(p=2)$ we have from eq. (\ref{cond}) that $\alpha^2=\beta^2$. This line corresponds to the $N=2$ supersymmetric sine-Gordon model \cite{11}. The field $\exp(a(\varphi_1+i\varphi_2))$ represents the lightest component of the superfield $\exp(aY(x,\theta))$. The calculation of the VEV (\ref{VEV2}) at this line gives the classical value $G(a,a,0)=1$ in exact agreement with the nonrenormalization theorem for $N=2$ supersymmetric QFT.

Some other tests for eq. (\ref{VEV2}) associated with QG restrictions of the QFT (\ref{a1}) will be given in the next section. Here we note that eqs. (\ref{VEV2}-\ref{Q_1}) simplify in the ``scaling'' limit $\alpha, \beta, \gamma \rightarrow \infty$, $a, b, c \rightarrow 0$ with $A, B, C$  fixed. This (strong coupling) limit corresponds to a weak coupling limit for the $\sigma$-model which describes the dual representation for the QFT (\ref{a1}) (see ref. \cite{11} for details). In the ``scaling'' limit with a fixed physical mass $M$ we obtain :
\begin{eqnarray}
G(a,b,c) &\simeq&\Big[ (\alpha^2)^{-\alpha^2} (\beta^2)^{\beta^2} (\gamma^2)^{\gamma^2}\Big]^{4(b^2+c^2-a^2)} \\
&&\times \Big(\frac{\gamma(\frac{1}{2}+P-2A)\gamma(\frac{1}{2}+P-2B)\gamma(\frac{1}{2}+P-2C)\cos 2\pi A}{\gamma(\frac{1}{2}+P)\cos 2\pi B\cos 2\pi C}\Big)^{\frac{1}{2}} \exp Q_1(A,B,C), 
\label{scaC} \nonumber
\end{eqnarray}
where $Q_1(A,B,C)$ is defined by eq. (\ref{Q_1}).

We intend to use eq. (\ref{scaC}) for the analysis of the VEVs of the observables in the $\sigma$-model describing the strong coupling limit of the QFT (\ref{a1}) in a separate publication. Here we shall discuss the application of eq. (\ref{VEV2}) to the perturbed CFT which can be obtained from QFT (\ref{a1}) by a QG restriction.

\section{Expectation values of local fields in parafermionic sine-Gordon theories}

\ \ \ \ For integer values of $n=4(\alpha^2-\gamma^2)=4\beta^2-2$ the QFT (\ref{a1}) admits a QG restriction with respect to the symmetry group $U_{q_1}(sl_2) (q_1 = \exp(\frac{i\pi}{2\beta^2}))$\,\footnote{$U_{q_1}(sl_2)$ is one of the symmetries of the QFT (\ref{a1}). Another symmetry group $U_{q_2}(sl_2) (q_2 = \exp(\frac{i\pi}{2\gamma^2}))$ \cite{11,14} is used in section 4 for the restriction of the QFT (\ref{a2}) to the integrable perturbed $SU(2)$ coset CFT models.} \cite{11}, giving the parafermionic sine-Gordon models \cite{15} with action : 
\begin{eqnarray}
{\cal{A}}_{n,\rho} = {\cal{A}}_n^{(0)} + \int d^2x \Big[\frac{1}{16\pi}(\partial_{\mu}\varphi)^2 - \kappa \Big( \psi \overline{\psi} e^{i\rho \varphi} + \psi^{\dagger} \overline{\psi}^{\dagger} e^{-i\rho \varphi}\Big)\Big],
\label{a2}
\end{eqnarray}
where $ {\cal{A}}_n^{(0)}$ is the action of the ${\mathbb{Z}}_n$ parafermionic CFT with a central charge $c = 2-\frac{6}{n+2}$ and the fields $\psi,\psi^{\dagger} (\overline{\psi}, \overline{\psi}^{\dagger})$ are the holomorphic (antiholomorphic) parafermionic currents with spin $\Delta = 1-\frac{1}{n}$ $(\overline{\Delta}=-\Delta)$. The field $\varphi$ is a scalar boson field and the parameter $\rho$ is given by :
\begin{eqnarray}
\rho^2 = \frac{p}{n(p+n)} = \frac{\gamma^2}{4\alpha^2(\alpha^2-\gamma^2)}.
\end{eqnarray}
The QFT (\ref{a2}) depends on two parameter $n$ and $p$ and we denote it by ${\cal{P}}(n,p)$.

Besides the conformal symmetry, the ${\mathbb{Z}}_n$ parafermionic CFT possesses an additional symmetry generated by the parafermionic currents $\psi (\overline{\psi})$. The basic fields in this CFT are the order parameters $\sigma_j, j=0,1,...,n-1$ with conformal dimensions :
\begin{eqnarray}
\delta_j = \frac{j(n-j)}{2n(n+2)}.
\label{delta}
\end{eqnarray}
All other fields in this CFT can be obtained from the fields $\sigma_j$ by application of the generators of the parafermionic symmetry \cite{18}. So it is natural to introduce as the basic operators in the QFT ${\cal{P}}(n,p)$ (\ref{a2}) the local fields :
\begin{eqnarray}
\Phi^{(j)}_{\omega} = \sigma_j \exp{(i\omega \varphi)},
\label{field2}
\end{eqnarray}
and to consider their VEVs.

For this we briefly describe the vacuum structure and the spectrum of the QFT ${\cal{P}}(n,p)$ (see ref. \cite{15} for details). ${\cal{P}}(n,p)$ possesses an explicit $U(1)$ symmetry. Besides this it has $n+1$ degenerate ground states $|0_s>$, $s=1,...,n+1$, which can be associated with the nodes of the Dynkin diagram $A_{n+1}$. The particles in ${\cal{P}}(n,p)$ are the kinks $A_{ss'}^{q}(\theta)$ characterized by the $U(1)$ charges $\ q=\pm 1$ and interpolating between vacua $s$ and $s'$ with $|s-s'|=1$. For $4\gamma^2=p<1$ there are also bound states of kinks $B_{ss'}^{(l)}(\theta)$ with $U(1)$ charge $q=0$ and $|s-s'|=0,2$. The masses $M_l$ of these particles are expressed through the mass $M$ of the kinks $A_{ss'}^{q}$ as :
\begin{eqnarray}
M_l = 2M\sin(\frac{\pi l p}{2}), \ \ \ \ l\leq \frac{1}{p}.\label{masses}
\end{eqnarray}
The relation between the parameter $\kappa$ in the action (\ref{a2}) and the mass $M$ of the basic kinks can be obtained by the Bethe Ansatz method \cite{19,20} and has the form :
\begin{eqnarray}
\frac{\pi \kappa}{n} \gamma(u) = \Big(\frac{\pi M \Gamma(\frac{p+n}{2})}{4\Gamma(\frac{p}{2})\Gamma(\frac{n+2}{2})} \Big)^{2u},\label{40}
\end{eqnarray}
where the function $\gamma$ is defined by eq. (\ref{gamma}) and 
\begin{eqnarray}
u = \frac{1}{n} - \rho^2 = \frac{1}{p+n}.\label{u}
\end{eqnarray}

To calculate the VEV $<0_s |\Phi_{\omega}^{(j)}| 0_s>$ we should express the fields (\ref{field2}) in terms of the fields $V(a,b,c)$ in QFT (\ref{a1}). For this purpose, we use the bosonization procedure for the fields $\sigma_j$ from the parafermionic CFT (see ref. \cite{21} for details). The fields $\sigma_j$ can be represented in terms of two boson fields $\varphi_2$ and $\eta$ with the stress-energy tensor :
\begin{eqnarray}
T_P = -\frac{1}{4}(\partial_z \varphi_2)^2 -\frac{1}{4}(\partial_z \eta)^2 + iq_{\beta}\partial^2_z\varphi_2,\label{tensor}
\end{eqnarray}
where $q_{\beta}$ is defined by the eq. (\ref{reflec1}).

The full stress-energy tensor including the field $\varphi$ has the form :
\begin{eqnarray}
T = T_P -\frac{1}{4}(\partial_z \varphi)^2,
\label{tens}
\end{eqnarray}
where the fields $\varphi$ and $\eta$ are related to the fields $\varphi_1$ and $\varphi_3$ from the restricted QFT (\ref{a1}) as :
\begin{eqnarray}
\eta(x) = \frac{2i}{\sqrt{n}}(-i\alpha \varphi_1 + \gamma \varphi_3)\ \ \ ,\ \ \ \varphi(x) = \frac{2}{\sqrt{n}}(-i\gamma \varphi_1 + \alpha \varphi_3).
\label{field3}
\end{eqnarray}
In the CFT with stress-energy tensor (\ref{tens}) these fields have the standard two point functions :
\begin{eqnarray}
<\eta(x)\eta(y)> = <\varphi(x)\varphi(y)> = -2\log|x-y|^2\ \ \ ,\ \ \ <\varphi(x)\eta(y)> =0 \nonumber
\end{eqnarray}
The fields $\sigma_j$ can be represented in terms of the fields $\varphi_2$ and $\eta$ as \cite{21} :
\begin{eqnarray}
\sigma_j = N_j^{-1} \exp(i\frac{j}{2\sqrt{n+2}}\varphi_2)\exp(\frac{j}{2\sqrt{n}}\eta).
\label{sigma_j}
\end{eqnarray}
Using eqs. (\ref{field3})(\ref{sigma_j}) we can express the operator $\Phi_{\omega}^{(j)}$ in terms of the field $V(a,b,c)$ as :
\begin{eqnarray}
\Phi_{\omega}^{(j)} = \sigma_j e^{i\omega \varphi} = N_j^{-1}V(a,b,c),
\label{field4}
\end{eqnarray}
where :
\begin{eqnarray}
\alpha a = A_{j,\omega} = \alpha^2(\frac{j}{n}+2\rho \omega)\ \ ,\ \  \gamma c = C_{j,\omega} = \gamma^2(\frac{j}{n}+\frac{2\omega}{\rho n})\ \ ,\ \ \beta b = B_{j,\omega} = A-C = \frac{j}{4}.
\label{coeff}
\end{eqnarray}
The numerical factor $N_j$ corresponds to the conformal normalization condition (\ref{correl}) for the fields $\Phi_{\omega}^{(j)}$. This factor can be expressed in terms of the ``reflection amplitude'' $R_{\beta}$ (\ref{reflec1}) associated to the reflection of the parameter standing before the twisted field $\varphi_2$ ($V(a,b,c) \rightarrow V(a,2q_{\beta}-b,c))$, by the relation \cite{7} :
\begin{eqnarray}
N_j^2 = \frac{R_{\beta}(a,b,c)}{R_{\beta}(0,0,0)} = \Big(\frac{\pi \mu}{4\beta^2}\Big)^{-\frac{2j}{n+2}}\frac{\gamma(\frac{j+1}{n+2})}{\gamma(\frac{1}{n+2})}.\label{48}
\end{eqnarray}
For the calculation of the VEV $<0_s | \Phi_{\omega}^{(j)} | 0_s>$ we should also take into account the factor $d_j(s)$ coming from the QG restriction of QFT (\ref{a1}). Following the conjecture of ref. \cite{7} we take it in the form :
\begin{eqnarray}
d_j(s) = \frac{\sin\big(\frac{\pi(j+1)s}{n+2}\big)}{\sin\big(\frac{\pi s}{n+2}\big)},
\label{d_j_s}
\end{eqnarray}
and using eqs. (\ref{field4}),(\ref{d_j_s}) we obtain :
\begin{eqnarray}
<0_s | \Phi_{\omega}^{(j)} | 0_s> = d_j(s) N_j^{-1} 
G(a,b,c) = d_j(s)  G_P(j,\omega) \nonumber
\end{eqnarray}
where
\begin{eqnarray}
G_P(j,\omega) &=& \Big(\frac{\pi M \Gamma(\frac{p+n+2}{2})}{2\Gamma(\frac{p}{2})\Gamma(\frac{n+2}{2})} \Big)^{2\omega^2+2\delta_j} \exp \Big(\int \frac{dt}{t} \Big[\Big( \frac{\sinh 4Ct \sinh 4(C-\gamma^2)t}{\sinh 2t \sinh 4\gamma^2t} \label{50}\\
&& \ \  - \frac{\sinh 4At \sinh 4(A-\alpha^2)t}{\sinh 2t \sinh 4\alpha^2t} - \frac{\sinh jt \sinh (n-j)t}{\tanh 2t \sinh(n+2)t}\Big) - 2(\omega^2 + \delta_j)e^{-2t}\Big]\Big),\nonumber 
\end{eqnarray} 
here $\delta_j$ is defined by eq. (\ref{delta}), and the dependence of $A_{j,\omega}$, $C_{j,\omega}$ on $j$ and $\omega$ is given by eq. (\ref{coeff}).

We consider in more detail some particular cases of eq. (\ref{50}). For arbitrary integer $n$ and $p=4\gamma^2$ we obtain 
\begin{eqnarray}
<e^{i\omega \varphi}> &=& G_P(0,\omega) = \Big(\frac{\pi M}{2n}\frac{\Gamma(\frac{p+n}{2})}{\Gamma(\frac{p}{2})\Gamma(\frac{n}{2})}\Big)^{2\omega^2}\exp Q_n(\omega), \nonumber
\end{eqnarray}
where
\begin{eqnarray}
Q_n(\omega) &=&  \int \frac{dt}{t} \Big( \frac{\sinh(nut)\sinh^2(2\rho\omega t)}{\sinh(2u t)\sinh t \sinh(\rho^2nt)} - 2\omega^2 e^{-2t}\Big), \label{G_P}
\end{eqnarray}
here and later $u$ is defined by eq. (\ref{u}).

For $n=1$, we have $\psi \equiv 1$ and the QFT ${\cal{P}}(1,p)$ coincides with the standard sine-Gordon model with a coupling constant $\rho^2 = \frac{p}{p+1}$. In this case, eq. (\ref{50}) gives exactly the VEV for the field $e^{i\omega \varphi}$ in the sine-Gordon theory proposed in \cite{4}.

For $n=2$ the parafermionic current $\psi(z)$ is a Majorana fermion and the theory ${\cal{P}}(2,p)$ describes an $N=1$ supersymmetric sine-Gordon theory with a coupling constant $\rho^2 = \frac{p}{2(p+2)}$. The field $\sigma \equiv \sigma_1$ is the spin field for the Majorana fermion $\psi(z)$ which has a square root branch point at the position of the field $\sigma$. As the result the theory has two sectors : the Neveu-Schwarz (NS) sector of local fields and the Ramond sector of non-local ones with respect to $\psi(z)$.

For $n=2$ we have three vacua $|0_s>, \ s=1,2,3$. This fact is in perfect agreement with the kink structure of the scattering theory for the $N=1$ supersymmetric sine-Gordon model. In the NS sector the VEVs of the fields $e^{i\omega \varphi}$ do not depend on $s$ and are described by eq. (\ref{G_P}) with $n=2$. In the Ramond sector the VEVs of the fields $\sigma e^{i\omega \varphi}$ are proportional to $\pm 1$ for $s=1,3$ and are equal to $0$ for $s=2$. They have the form :
\begin{eqnarray}
<\sigma e^{i\omega \varphi}> &=& (\pm 1,0)\sqrt{2} \Big( \frac{\pi M p}{8}\Big)^{\frac{1}{8}+2\omega^2} \exp Q_R(\omega),\label{G_R}
\end{eqnarray} 
where
\begin{eqnarray}
Q_R(\omega) &=&  \int \frac{dt}{t} \Big[\frac{1}{\cosh ut} \Big(\frac{1}{4\cosh \rho^2t \cosh \frac{t}{2}} - \frac{1}{8\cosh ut}\nonumber \\
 &+& \frac{\cosh(1-u)t \sinh^2 2\rho \omega t}{\sinh t \sinh(2\rho^2t)}\Big) - (\frac{1}{8} + 2\omega^2) e^{-2t}\Big]. \nonumber
\end{eqnarray}

Another particular case that we consider here corresponds to arbitrary $n$ and $p = 4\gamma^2 = 2$. The QFT  ${\cal{P}}(n,2)$ posseses $N=2$ supersymmetry and is known as the $N=2$ supersymmetric restricted sine-Gordon model. The $N=2$ superconformal algebra with Virasoro central charge 
\begin{eqnarray}
c_n = \frac{3n}{n+2} \label{cent}
\end{eqnarray}
posseses a representation in terms of a ${\mathbb{Z}}_n$ parafermionic CFT and a massless scalar field $\varphi(z)$ with stress-energy tensor (\ref{tensor}). The supersymmetric currents $S(z)$, $S^{\dagger}(z)$ and the $U(1)$ current $J(z)$ in this representation \cite{22} have the forms :
\begin{eqnarray}
S(z) &=& \big(\frac{2c_n}{3} \big)^{\frac{1}{2}} \psi \exp(i\nu_n \varphi)\ \ ,\ \ S^{\dagger}(z) = \big(\frac{2c_n}{3} \big)^{\frac{1}{2}} \psi^{\dagger} \exp(-i\nu_n \varphi), \label{54} \\
&&\ \ \ \ \ \ \ \ \ \ \ \ \ \ \ \ \ J(z) = \big(\frac{c_n}{12}\big)^{\frac{1}{2}} \partial_z \varphi, \nonumber
\end{eqnarray}
where $\nu_n = (\frac{n+2}{2n})^{\frac{1}{2}}$ and $(\psi(z)$,$\psi^{\dagger}(z))$ are the parafermionic currents with spins $\Delta = 1 - \frac{1}{n}$.

All the fields in the minimal $N=2$ CFT with the central charge (\ref{cent}) can be expressed in terms of fields from the ${\mathbb{Z}}_n$ CFT and the field $\varphi$. In particular, the so called ``chiral'' fields $ {\chi}_j, \ j=0,1,...,n$, which satisfy the condition
\begin{eqnarray}
G_{-\frac{1}{2}}^{\dagger} {\chi}_j = 0,
\end{eqnarray}
can be represented as :
\begin{eqnarray}
\chi_j(x) = \sigma_j(x)\exp\big(-i j \nu_n \varphi(x)/(n+2)\big), \label{56}
\end{eqnarray}
and have conformal dimensions $\Delta_j = \frac{j}{2(n+2)}$. These fields, normalized by the condition (\ref{correl}) have the operator product expansion 
\begin{eqnarray}
\chi_j(x) \chi_l(0) = C_{jl} \chi_{j+l}(0) + O(|x|^\sigma);\ \ \sigma > 0,
\end{eqnarray}
with structure constants $C_{jl}$ which have the factorized form \cite{18} :
\begin{eqnarray}
C_{jl} = \frac{a_j a_l}{a_{j+l}};\ \ a_j = \Big(\frac{\gamma(\frac{1}{n+2})}{\gamma(\frac{j+1}{n+2})}\Big)^{\frac{1}{2}}\label{58}
\end{eqnarray}

For the analysis of an $N=2$ CFT it is more convinient to work with renormalized chiral fields $\phi_j$ : 
\begin{eqnarray}
\phi_j = \frac{\chi_j}{a_j} ;\ \  j=0,1,...,n,
\end{eqnarray}
which have the simple operator algebra :
\begin{eqnarray}
\phi_i(x) \phi_j(0) = \phi_{i+j}(0) + O(|x|^\sigma);\ \ \sigma > 0,
\end{eqnarray}
and play the roles of order parameters in the perturbed CFT ${\cal{P}}(n,2)$. The expectation values of the fields $\phi_j$ can be calculated using eqs. (\ref{50}),(\ref{56}),(\ref{58}) and have the simple form :
\begin{eqnarray}
<\phi_j> = <0_s| \phi_j |0_s> = \Big(\frac{\pi M}{4}\Big)^{\frac{j}{k+2}} \frac{\sin\big(\frac{\pi (j+1)s}{k+2}\big)}{\sin\big(\frac{\pi s}{k+2}\big)}.
\end{eqnarray}

Using the notation $t = \Big(\frac{\pi M}{4}\Big)^{\frac{1}{k+2}}$ and $X = <\phi_1>$. Then we can express the VEVs of the fields $\phi_j$ in terms of $X$ :
\begin{eqnarray}
<\phi_j> = t^jT_j(\frac{X}{2t}),
\end{eqnarray}
where $T_j(z)$ is the Chebychev polynomial of the second type defined by the relations :
\begin{eqnarray}
T_j(z) = \frac{\sin(j+1)y}{\sin y} ; \ \ \cos y = z \label{T_j}
\end{eqnarray}
In particular, from the condition $\phi_{n+1}=0$ we obtain the equation for the VEV of the order parameter $\phi_1$ :
\begin{eqnarray} 
t^{n+1}T_{n+1}(\frac{X}{2t}) = 0.
\end{eqnarray}
The equation (\ref{T_j}) defines the critical points of the Landau-Ginsburg superpotential 
\begin{eqnarray}
W(x) = \frac{2 t^{n+1}}{n+2} U_{n+2}(\frac{X}{2t}),\label{W}
\end{eqnarray}
where $U_{n+2}(z)$ is the Chebychev polynomial of the first type defined by the relations $U_n(z) = \cos ny, \ \cos y = z; \ U'_n(z) = nT_{n-1}(z)$ . The function (\ref{W}) coincides exactly with the Landau-Ginsburg superpotential for the perturbed $N=2$ minimal CFT conjectured in ref. \cite{23}.

At the end of this section we consider briefly the VEVs in the QFT described by the action 
\begin{eqnarray}
{\cal{A}}_{n} = {\cal{A}}_n^{(0)} - \kappa \int d^2x \Big( \psi \overline{\psi} + \psi^{\dagger} \overline{\psi}^{\dagger}\Big),
\label{a3}
\end{eqnarray}
This integrable QFT was introduced and studied in refs. \cite{24},\cite{25}, where its scattering theory and thermodynamics were described. The most interesting feature of this QFT is related to its large $n$ limit where it describes an $O(3)$-nonlinear $\sigma$ model.

The VEVs of the fields $\sigma_j$ in the QFT (\ref{a3}) can be obtained by taking properly limit $\rho \rightarrow 0, \omega \rightarrow 0$ from eq. (\ref{50}). The limit $\rho \rightarrow 0$ corresponds to $4\gamma^2 = p \rightarrow 0$. As it was noted in the beginning of this section for $p<1$, the QFT ${\cal{P}}(n,p)$ contains bound states $B^{(l)}_{ss'}$ of the basic kinks $A_{ss'}$, with masses (\ref{masses}). The QFT (\ref{a3}) can be obtained from the QFT (\ref{a2}) if we fix the parameter $m \equiv m_1$. Then in the limit $p \rightarrow 0$ the kinks $A_{ss'}$ have infinite mass and the only stable particles that survive after this limit are the kinks  $B_{ss'}\equiv B^{(1)}_{ss'}$. The relation between the mass $m$ of these kinks and parameter $\kappa$ in the action (\ref{a3}) can be obtained from eqs. (\ref{masses}),(\ref{40}) and has the form \cite{25} :
\begin{eqnarray}
\pi \kappa \gamma(1/n) = n\Big(\frac{m}{4n}\Big)^{\frac{2}{n}}.\label{67}
\end{eqnarray}

The kinks $B_{ss'}$ interpolate between the vacua $s$ and $s'$ with $|s-s'|=0,2$.  The subspace of the ground states $|0_s>$ with odd $s=1,3,...\leq n+1$
is invariant under the dynamics of (\ref{a3}) and form the space of vacua for this QFT. Taking the limit $p=4\gamma^2 \rightarrow 0$, $m=$ fixed in eq. (\ref{50}), we obtain :
\begin{eqnarray}
<0_s |\sigma_j| 0_s> = \frac{\sin\big(\frac{\pi(j+1)s}{n+2}\big)}{\sin\big(\frac{\pi s}{n+2}\big)} \Big(\frac{M}{4}\Big)^{2\delta_j}\exp Q_j,\label{68}
\end{eqnarray}
where
\begin{eqnarray}
Q_j = \int \frac{dt}{t} \Big( \frac{\sinh(tj)\sinh((n-j)t)}{\tanh(nt)\sinh((n+2)t)} - 2\delta_j e^{-2t}\Big)\nonumber.
\end{eqnarray}
In the limit $n \rightarrow \infty$, $j\ll n$, eq. (\ref{68}) simplifies. In this limit we can consider $\pi s /(n+2)$ as continuous variable $0< {\vartheta} <\pi$. The integral $Q_j$ in this limit also can be calculated and we obtain : 
\begin{eqnarray}
<\sigma_j>_{\vartheta} = (j+1)^{\frac{1}{2}}\chi_{j/2}(2{\vartheta}) + {\cal{O}}(\frac{1}{n}),\label{69}
\end{eqnarray}
where $\chi_{j/2}({\vartheta})$ are the characters of the representations of $SU(2)$ with spins $j/2$.

\section{Vacuum expectation values of local fields in the integrable perturbed $SU(2)$ coset model}
For integer values of $p = 4\gamma^2$ (in this case $4\alpha^2 = p+n$ is also integer) the QFT ${\cal{P}}(n,p)$ admits an additional restriction with respect to the QG $U_{q_2}(sl_2) (q_2=\exp(\frac{2i\pi}{p}))$ to give the $SU(2)$ coset CFT models \cite{16} perturbed by an integrable operator $\Phi_C$ with conformal dimension $\Delta = 1  -\frac{2}{p+n}$ \cite{15}. The actions of these QFTs, which we denote as ${\mathfrak{C}}(n,p)$, have the form :
\begin{eqnarray}
{\cal{A}}_{n,p} = {\cal{A}}_{n,p}^{(0)} - \lambda \int d^2x \Phi_C(x),\label{70}
\label{a4}
\end{eqnarray}
where ${\cal{A}}_{n,p}^{(0)}$ is the action of the $\frac{SU_2(n)\times SU_2(p-2)}{SU_2(n+p-2)}$ coset CFT ${\mathbb{C}}(n,p)$ with a central charge :
\begin{eqnarray}
c_{n,p} = \frac{3n(p-2)}{n+p}\Big(\frac{1}{n+2} +\frac{1}{p}\Big).\label{71}
\end{eqnarray}
The stress-energy tensor of the ${\mathbb{C}}(n,p)$ models can be obtained from the stress-energy tensor (\ref{tens}) by twisting the field $\varphi$, namely :
\begin{eqnarray}
T_C = T_P - \frac{1}{4}(\partial_z\varphi)^2 + iq_{\rho}\partial_z^2\varphi,\label{72} \ \ \mbox{where} \ \ 2q_{\rho} = \rho - 1/n\rho= - u/\rho.
\end{eqnarray}
The conformal dimensions of the basic fields $\Phi^{(j)}_{lm}$ in the ${\mathbb{C}}(n,p)$ models are characterized by three integers $j,l,m$ and have the form :
\begin{eqnarray}
\Delta^{(j)}_{lm} = \frac{(pm-(p+n)l)^2-n^2}{4n(p+n)} + \frac{j(n-j)}{2n(n+2)}\label{73}
\end{eqnarray}
where the integers $l<p$ and $m<p+n$ satisfy the condition 
\begin{eqnarray}
|m-l-kn|=j ; \ \ k \in {\mathbb{Z}}_n.\label{74}
\end{eqnarray}
The fields $\Phi^{(j)}_{lm}$ can be represented in terms of fields $\sigma_j$ from the ${\mathbb{Z}}_n$ CFT and the field $\varphi$ as \cite{22} :
\begin{eqnarray}
\Phi^{(j)}_{l,m} = N^{-1}_{j,lm}\sigma_j e^{i\omega_{lm}\varphi},\label{75}
\end{eqnarray}
where 
\begin{eqnarray}
2\omega_{lm}=-\rho(m-1)+(l-1)/n\rho; \ \ \ \rho^2=\frac{p}{n(p+n)}.\nonumber
\end{eqnarray}
The normalization factor $N_{j,lm}$ corresponds to the conformal normalization condition (\ref{correl}). It can be calculated using the integral representation for the correlation functions of the ${\mathbb{C}}(n,p)$ models with screening charges :
\begin{eqnarray}
Q_1=\kappa \int d^2x \psi \overline{\psi}e^{i\rho\varphi}\ \ \ ; \ \ \ Q_2=\kappa' \int d^2x \psi^{\dagger} \overline{\psi}^{\dagger}e^{-i\frac{1}{n\rho}\varphi}\nonumber
\end{eqnarray}
The same result can be obtained if we express $N_{j,lm}$ using the relation (\ref{48}) through the ``reflection amplitude'' $R_{\rho}(j,\omega)$ associated with the reflection $\omega \rightarrow 2q_{\rho}-\omega$. Namely :
\begin{eqnarray}
N_{j,lm}^2 = \frac{R_{\rho}(j,\omega_{lm})}{R_{\rho}(0,0)},
\end{eqnarray}
where the function $R_{\rho}(j,\omega)$ is defined by the equation ;
\begin{eqnarray}
G_P(j,\omega)  = R_{\rho}(j,\omega) G_P(n-j,2q_{\rho}-\omega)
\end{eqnarray}
and has the form : 
\begin{eqnarray}
R_{\rho}(j,\omega) &=& \Big(\frac{\pi \kappa \gamma(u)}{n(\rho^4 n^2)^{1/n}} \Big)^{-\frac{2\overline{\omega}}{\rho}}(\rho^2 n)^{\frac{(2j-n)}{n}} \label{78} \nonumber \\
&&\times \frac{\Gamma(\frac{1}{2}-\frac{(2j-n)}{2n}-2\rho\overline{\omega})}{\Gamma(\frac{1}{2}-\frac{(2j-n)}{2n}+2\rho\overline{\omega})} \frac{\Gamma(\frac{1}{2}+\frac{(2j-n)}{2n}+2\overline{\omega}/\rho n)}{\Gamma(\frac{1}{2}+\frac{(2j-n)}{2n}-2\overline{\omega}/\rho n)}
\end{eqnarray}
where \ \ \ \ \ \  $\overline{\omega} = \omega -q_{\rho}$.

The models ${\mathfrak{C}}(n,p)$ possess a discrete set of degenerate ground states $|0_{sr}>$ which are labelled by two integers $s=1,...,n+1$ and $r=1,...,p-1$. Their spectrum consists of the kinks $A_{ss'}^{rr'}$ interpolating between the vacua $|0_{sr}>$ and $|0_{s'r'}>$ with $|s-s'|=|r-r'|=1$. The scattering theory of these excitations is described in \cite{15}. The mass $M$ of the kinks is connected to the parameter $\lambda$ in action (\ref{a4}) by the relation \cite{20} :
\begin{eqnarray}
\pi\lambda\big(\gamma(u)\gamma(3u)\big)^{\frac{1}{2}} = \frac{n(p-2)}{n+p-2}\Big[ \frac{\pi M}{4} \frac{\Gamma(\frac{p+n+2}{2})}{\Gamma(\frac{p}{2})\Gamma(\frac{n+2}{2})}\Big]^{4u}\label{79}
\end{eqnarray}

To calculate the VEVs of the fields $\Phi^{(j)}_{lm}$ we can use eqs. (\ref{75}),(\ref{50}) and take into account the factor that comes from the QG restriction of QFT ${\cal{P}}(n,p)$. This factor, conjectured in \cite{7}, has the form :
\begin{eqnarray}
\tilde{d}_{lm}(r)=(-1)^{k(r-1)} \frac{\sin\big(\frac{\pi l r}{p}\big)}{\sin\big(\frac{\pi r}{p}\big)}
\end{eqnarray}
where integers $k$ and $l$ are defined by eqs. (\ref{73}), (\ref{74}).

As the result we obtain 
\begin{eqnarray}
<\Phi^{(j)}_{l,m}> &=& d_j(s)\tilde{d}_{lm}(r)N^{-1}_{j,lm}G_P(j,\omega_{lm}) = d_j(s) \tilde{d}_{lm}(r)G_C(j,\omega_{lm}),\label{81}
\end{eqnarray}
where 
\begin{eqnarray}
G_C(j,\omega) &=& \Big[\frac{\pi M}{2} \frac{\Gamma(\frac{p+n+2}{2})}{\Gamma(\frac{p}{2})\Gamma(\frac{n+2}{2})} \Big]^{2\Delta^{(j)}_{lm}} \exp \int \frac{dt}{t} \Big[ \frac{\sinh(4Ct)\sinh(4(C-\gamma^2+\frac{1}{2})t)}{\tanh 2t \sinh pt}\nonumber \\
&-&\frac{\sinh(4At)\sinh(4(A-\alpha^2+\frac{1}{2})t)}{\tanh 2t \sinh(p+n)t} -
\frac{\sinh(jt)\sinh(n-j)t)}{\tanh 2t \sinh(n+2)t} - 2\Delta^{(j)}_{lm}e^{-2t}\Big],\nonumber
\end{eqnarray}
here $\Delta^{(j)}_{lm}$ are defined by eq. (\ref{73}), and the dependence of $A_{j,\omega}$, $C_{j,\omega}$ on $j$ and $\omega$ is given by eq. (\ref{coeff}).

We consider in more detail several particular cases of eq. (\ref{81}). For arbitrary $n,p$ and $j=0$ we obtain :
\begin{eqnarray}
<\Phi_{lm}^{(0)}> &=& \Big[\frac{\pi M}{2n} \frac{\Gamma(\frac{p+n}{2})}{\Gamma(\frac{p}{2})\Gamma(\frac{n}{2})} \Big]^{2\Delta^{(0)}_{lm}} \tilde{d}_{lm}(r) \exp Q_{n,p}(\omega_{lm}),\label{82}
\end{eqnarray}
where
\begin{eqnarray}
Q_{n,p}(\omega) &=& \int \frac{dt}{t} \Big[\frac{\sinh(nut)\sinh(2\rho\omega t)\sinh(2(\rho \omega+u)t)}{\tanh 2ut \sinh t \sinh (n\rho^2t)} - 2\Delta^{(0)}_{lm} e^{-2t}\Big].\nonumber
\end{eqnarray}

For $n=1$ the QFT ${\mathfrak{C}}(1,p)$ describes minimal CFT models perturbed by the operator $\Phi_{13}$. In this case eq. (\ref{82}) coincides  exactly with the expression for the VEVs of primary fields in the perturbed minimal CFT proposed in \cite{7}.

For $n=2$ the QFT ${\mathfrak{C}}(2,p)$ describes N=1 supersymmetric minimal CFT models perturbed by the operator $\Phi_C$, preserving supersymmetry. These theories possess two sectors. The fields local with respect to supercurrent $S(z)$ form the NS sector. The primary fields in this sector are characterized by $j=0$. Their VEVs are given by eq. (\ref{82}) with $n=2\ (\rho^2=\frac{p}{2(p+2)}, u=\frac{1}{p+2})$. In particular the VEV of the field $\Phi_{13}^{(0)}$, which is the superpartner of the operator $\Phi_C$, has the rather simple form :
\begin{eqnarray}
<\Phi_{13}^{(0)}>= \pm \Big(\frac{\pi M p}{8} \Big)^{1-4u}\Big( \gamma(u)\gamma(3u)\Big)^{\frac{1}{2}}\label{83}
\end{eqnarray}
The fields non-local with respect to $S$ form the Ramond sector. The primary fields there are characterized by $j=1$. The VEVs of these fields can be represented in the form :
\begin{eqnarray}
<\Phi_{lm}^{(R)}>= \sqrt{2}(\pm 1,0)\tilde{d}_{lm}(r)\Big(\frac{\pi M p}{8}\Big)^{2\Delta^{(1)}_{lm}}\exp Q_{R,p}(\omega_{lm}),\label{84}
\end{eqnarray}
where
\begin{eqnarray}
Q_{R,p}(\omega) &=& \int \frac{dt}{t} \Big[ \frac{\cosh 2ut}{\cosh ut} \Big(-\frac{1}{8\cosh ut \cosh 2ut} +\frac{\cosh 2ut}{4\cosh \rho^2t \cosh \frac{t}{2}} \nonumber \\
 &+& \frac{\cosh(1-u)t\sinh2\rho \omega t \sinh2(\rho \omega+u)t}{\sinh t \sinh 2\rho^2 t } \Big) -2\Delta^{(1)}_{lm}e^{-2t}\Big].\nonumber
\end{eqnarray}

The first supersymmetric model ${\mathfrak{C}}(2,3)$ describes the tricritical Ising model perturbed by a subleading thermal operator with dimension $\Delta_T = \frac{3}{5}$. This model coincides with the perturbed minimal model ${\mathfrak{C}}(1,4)$. One can check that VEVs of the fields calculated with eqs. (\ref{84}) and (\ref{82}) (with $n=2$, $\rho^2=\frac{3}{10}$, $u=\frac{1}{5}$) coincide with the VEVs of the same fields calculated from eq. (\ref{82}) (with $n=1$, $\rho^2=\frac{4}{5}$, $u=\frac{1}{5}$).

The next model ${\mathfrak{C}}(2,4)$ possesses $N=2$ supersymmetry. The CFT ${\mathbb{C}}(2,4)$ is characterized by a central charge c=1. Its action coincides with the action of a massless free scalar field $\varphi$. The representation of the $N=2$ superalgebra in terms of the field $\varphi$ corresponds to the case $n=1$ in eq. (\ref{54}). The CFT ${\mathbb{C}}(2,4)$ describes the special point at the critical line of the Ashkin-Teller (${\mathbb{Z}}_4$) model. This critical line can be parametrized by the conformal dimension $\Delta_\epsilon=\beta^2$ of the thermal operator $\epsilon=\sqrt{2}\cos(\beta\varphi)$. At the $N=2$ supersymmetric point $\beta^2=\frac{2}{3}$ the thermal operator $\epsilon$ coincides with operator $\Phi_C$. The action of the Ashkin-Teller model perturbed by the thermal operator has the form of a sine-Gordon model :
\begin{eqnarray}
{\cal{A}} = \int d^2x \Big[\frac{1}{16\pi}(\partial_{\mu}\varphi)^2 - \sqrt{2}\lambda\cos(\beta \varphi)\Big]. \label{85}
\end{eqnarray}
At the point $\beta^2=\frac{2}{3}$ corresponding to the ${\mathbb{C}}(2,4)$ model it possesses $N=2$ supersymmetry.

The order parameters in the Ashkin-Teller model are the fields $\sigma,\sigma^{\dagger}$ and the field $\Sigma \sim \sigma^2$. The field $\sigma$  has conformal dimension $\Delta_\sigma=\frac{1}{16}$ which is independent of $\beta$ along the whole critical line. This field is non-local with respect to the field $\varphi$ which has the square-root branch point at the position of $\sigma$. The field $\Sigma$ is local with respect to $\varphi$. It has conformal dimension $\Delta_{\Sigma}=\beta^2/4$ and can be represented in terms of $\varphi$ as $\Sigma = \pm \sqrt{2}\cos(\beta \varphi /2)$. At $\beta^2 = \frac{2}{3}$, $\Delta_{\Sigma}=\frac{1}{6}$ the field $\Sigma$ coincides with the NS field $\Phi_{13}^{(0)}\in{\mathbb{C}}(2,4)$ which has the same dimension. The VEV of $\Sigma$ can be calculated independently using eq. (\ref{83}) with $p=4, u=\frac{1}{6}$ and also from the VEV for the field $\cos(\beta \varphi /2)$ in sine-Gordon model (\ref{85}). Both results are in exact agreement : 
\begin{eqnarray}
<\Sigma> = \pm \Big(\frac{\pi M}{2} \Big)^{\frac{1}{3}}\Big(\gamma(1/6)\Big)^{\frac{1}{2}} = \pm \sqrt{2}<\cos(\beta \varphi /2)>_{\beta^2=\frac{2}{3}}.
\end{eqnarray}
The normalized Ramond field $\Phi_{12}^{(R)}\in{\mathbb{C}}(2,4)$ has dimension $\Delta_{12}^{(1)}=\frac{1}{16}$ and can be represented in terms of $\sigma$ as : $\Phi_{12}^{(R)} = 2^{-\frac{1}{2}}(\sigma+\sigma^{\dagger})$. The VEV of the field $\sigma+\sigma^{\dagger}$ can be calculated from eq. (\ref{84}) and we obtain : 
\begin{eqnarray}
<\sigma+\sigma^{\dagger}> = 2(\pm1,0)(\pi M)^{\frac{1}{8}} \exp\int\frac{dt}{8t}\Big(\frac{1}{\cosh t \cosh t/3} - e^{-2t}\Big).
\end{eqnarray}
This result is also in exact agreement (for $\beta^2=\frac{2}{3}$) with the VEV for the non-local spin field $\sigma$ in the sine-Gordon (perturbed Ashkin-Teller) model with coupling constant $\beta$, which has the form :
\begin{eqnarray}
<\sigma+\sigma^{\dagger}>_{\beta} &=& 2(\pm 1,0)\Big( \frac{2M\sqrt{\pi}\Gamma(\frac{1}{2(1-\beta^2)})}{\Gamma(\frac{\beta^2}{2(1-\beta^2)})}\Big)^{\frac{1}{8}} \nonumber \\
\ \ \ \ \ \ \ \ && \times \ \exp \int \frac{dt}{8t}\Big(\frac{\cosh2(1-\beta^2)t}{\cosh\beta^2t \cosh(1-\beta^2)t \cosh t} - e^{-2t}\Big).\label{88}
\end{eqnarray}

Some other applications of eq. (\ref{81}) as well as the derivation of eq. (\ref{88}) shall be given in another publication.

\section{Concluding remarks}
\ \ \ \ In the previous sections we calculated VEVs in perturbed CFT using the data of the corresponding conformal theory. In this section, which is not directly related to this subject, we illustrate the application of the VEVs in a massive theory to the calculation of some quantities in a basic CFT. As an example, we consider a theory which can be obtained by the analytical continuation ($\rho \rightarrow b/i$) from the QFT (\ref{a2}) (parafermionic sinh-Gordon model). This model is described by the action :
\begin{eqnarray}
{\cal{A}}_{n,b} = {\cal{A}}_n^{(0)} + \int d^2x \Big[\frac{1}{16\pi}(\partial_{\mu}\varphi)^2 - \kappa \Big( \psi \overline{\psi} e^{b\varphi} + \psi^{\dagger} \overline{\psi}^{\dagger} e^{-b\varphi}\Big)\Big],
\label{89}
\end{eqnarray}
where ${\cal{A}}_n^{(0)}$ is the action of the ${\mathbb{Z}}_n$ parafermionic CFT.

For simplicity we consider only the exponential fields $V_a(x)=\exp(a\varphi(x))$ in the QFT (\ref{89}). The VEVs of these fields expressed in terms of the parameter $\kappa$ can be obtained by an analytical continuation ($\rho \rightarrow b/i$, $\omega \rightarrow a/i$) from eqs. (\ref{G_P})(\ref{40}) and have the form :
\begin{eqnarray}
<V_a>=G_n(a)=\Big(\frac{\pi\kappa\gamma(v)}{n} \Big)^{-a^2/v}\exp\Big(-\int\frac{dt}{t}\Big[\frac{\sinh nvt\sinh^2(2bat)}{\sinh2vt \sinh t \sinh nb^2t} - 2a^2e^{-2t}\Big] \Big),\label{90}
\end{eqnarray}
here and later $v=1/n+b^2$.

The QFT (\ref{89}) can be defined as a perturbed CFT where the first three terms in (\ref{89}) define the action ${\cal{A}}_{n,b}^{(0)}$ of the CFT and the last one the perturbation :
\begin{eqnarray}
{\cal{A}}_{n,b}^{(0)}={\cal{A}}_{n}^{(0)} + \int \Big[\frac{1}{16\pi}(\partial_{\mu}\varphi)^2 - \kappa \psi \overline{\psi} e^{b\varphi} \Big].\label{91}
\end{eqnarray}
For $n=1,2$ this CFT coincides with the Liouville and supersymmetric Liouville models. It has the stress-energy tensor :
\begin{eqnarray}
T(z) = T_P -\frac{1}{4}(\partial_z \varphi)^2 + Q\partial_z^2\varphi,
\label{92}
\end{eqnarray}
where $T_P$ is stress-energy tensor of the ${\mathbb{Z}}_n$ parafermionic CFT and $2Q = b + 1/nb = v/b$.

The fields $V_a = e^{a\varphi}$ are the primary fields for the CFT (\ref{91}). The ``reflection amplitudes'' for these fields can be obtained from the ``reflection relations'' for the VEVs (\ref{90}):
\begin{eqnarray}
G_n(a) = R_b^{(n)}(a)G_n(2Q-a).
\end{eqnarray}
These can be calculated from eq. (\ref{90}) or by analytical continuation from eq. (\ref{78}) (with $j=0$) and have the form :
\begin{eqnarray}
R_b^{(n)}(a) = \Big( \frac{\pi\kappa\gamma(v)}{n(n^2b^4)^{1/n}}\Big)^{-2\overline{a}/b}\frac{\Gamma(1+2b\overline{a})\Gamma(1+2\overline{a}/nb)}{\Gamma(1-2b\overline{a})\Gamma(1-2\overline{a}/nb)}, \label{94}
\end{eqnarray}
where \ \ \ \ \ \ $\overline{a}=a-Q$.

We consider the three point functions (structure constants) of the fields $V_a$ in the CFT (\ref{91}) :
\begin{eqnarray}
{\mathfrak{J}}^{(n)}(a_1,a_2,a_3) = <V_{a_1}(x_1)V_{a_2}(x_2)V_{a_3}(x_3)>\label{95}
\end{eqnarray}
To remove the trivial dependence on $x_i$ we put $x_1=0$, $x_2=(1,0)$, $x_3=\infty$. The function (\ref{95}) satisfies the ``reflection relations'' with amplitude $R_b^{(n)}$ under all the transformations $a_i \rightarrow 2Q - a_i$. For $n=1$ and $2$ this function was proposed in \cite{6},\cite{27}. To construct a function satisfying proper reflection properties, we introduce functions $Y_n(a)$ and $\Upsilon_n(a)$ related to the VEVs $G_n(a)$ by :
\begin{eqnarray}
Y_n(a) = \frac{G_n(a/2)G_n(Q - a/2)}{G_n^2(Q/2)} = \Big( \frac{\pi\kappa\gamma(v)}{n}\Big)^{-\frac{(a-Q)}{2v}}\Upsilon_n(a),\label{96}
\end{eqnarray}
\begin{eqnarray}
\Upsilon_n(a) = \exp \Big(-\int \frac{dt}{t}\Big[ \frac{\sinh(nvt)\sinh^2b(a-Q)t}{\sinh vt \sinh t \sinh nb^2t } -(a-Q)^2 e^{-2t}\Big]\Big)\nonumber
\end{eqnarray}
These functions satisfy the properties :
\begin{eqnarray}
Y_n(a)=Y_n(2Q-a); \ \ Y_n(0)=0; \ \ Y_n(Q)=1; \ \ Y_n(2(2Q-a))=R_b^{(n)}(a)Y_n(2a);\nonumber
\end{eqnarray}
\begin{eqnarray}
\Upsilon_n(a+nb) &=& \prod_{j=0}^{n-1}\gamma(ba+jv) \Upsilon_n(a);\label{97}\\
\Upsilon_n(a+1/b) &=& (nb^2)^{\frac{2ab+1-v}{b^2}}\prod_{j=0}^{n-1}\gamma(\frac{a}{nb}+\frac{jv}{nb^2})\Upsilon_n(a)\nonumber
\end{eqnarray}
The function (\ref{95}), which satisfies the reflection properties, coincides with the solutions for $n=1,2$ and possesses similar factorization properties, can be written in the form :\begin{eqnarray}
{\mathfrak{J}}^{(n)}(a_1,a_2,a_3) &=& \frac{{\dot{Y}}_n(0)Y_n(2a_1)Y_n(2a_2)Y_n(2a_3)}{Y_n(p-2Q)Y_n(p-2a_1)Y_n(p-2a_2)Y_n(p-2a_3)}\label{98} \\
&=& \Big(\frac{\pi\kappa\gamma(v)}{n}\Big)^{(2Q-p)/b}\frac{{\dot{\Upsilon}}_n(0)\Upsilon_n(2a_1)\Upsilon_n(2a_2)\Upsilon_n(2a_3)}{\Upsilon_n(p-2Q)\Upsilon_n(p-2a_1)\Upsilon_n(p-2a_2)\Upsilon_n(p-2a_3)},\nonumber
\end{eqnarray}
where $p=a_1+a_2+a_3$ and ${\dot{Y}}_n(0) = \frac{dY_n}{da}(0)$,\ \ \ 
${\dot{\Upsilon}}_n(0) = \frac{d\Upsilon_n}{da}(0)$.

The expression (\ref{98}) possesses the correct analytical properties and semi-classical ($b<<1$) behaviour, which is governed by the Liouville-like counterterm $\tilde{\kappa}e^{nb\varphi}$ regularizing the action (\ref{91}),
\begin{eqnarray}
{\mathfrak{J}}_{\kappa,b}^{(n)}(a_1,a_2,a_3) \simeq {\mathfrak{J}}_{\tilde{\kappa},nb}^{(1)}(a_1,a_2,a_3);\ \ \ \ \tilde{\kappa} \simeq -\frac{\pi}{b^2}\Big(\frac{\pi\kappa\gamma(\frac{1}{n})}{n}\Big)^n.\nonumber
\end{eqnarray}
We propose eq. (\ref{98}) as an exact three-point correlation function in the CFT (\ref{91}).

As a function of $\ \ p=a_1+a_2+a_3\ \ $\  \ ${\mathfrak{J}}^{(n)}$ possesses poles at $\ p_{k,l}=2Q-kb-l/nb, \ \ k-l=0\ mod\ n$ (and hence, due to (\ref{97}), at\ $\ \tilde{p}_{k,l} = 4Q+kb+l/nb$). In particular, the residue at the pole\ $\ p_{mn,0}=2Q-nmb$ coincides with the term of order $\kappa^{nm}$ in the perturbative expansion for ${\mathfrak{J}}^{(n)}(a_1,a_2,a_3)$ with action (\ref{91}) :
\begin{eqnarray}
\raisebox{-0.4cm}{~\shortstack{${Res\ {\mathfrak{J}}^{(n)}(a_1,a_2,a_3)}$\\${p=2Q-nmb}$}} = \frac{\kappa^{nm}}{(nm)!}\langle V_{a_1}(x_1)V_{a_2}(x_2)V_{a_3}(x_3)\Big(\int\psi\overline{\psi}e^{b\varphi(u)}d^2u\Big)^{nm}\rangle\nonumber.
\end{eqnarray}
This residue can be calculated using the functional relations (\ref{97}). In this way we arrive at the following expression for the parafermionic generalization of the integral calculated in \cite{17} :
\begin{eqnarray}
&&\frac{1}{(nm)!} \int \langle V_{a_1}(x_1)V_{a_2}(x_2)V_{a_3}(x_3) \prod_{s=1}^{nm}\psi(u_s)\overline{\psi}(u_s)e^{b\varphi(u_s)}\rangle du_1...du_{nm}\label{99} \\ &=& \Big( \frac{\pi\gamma(v)}{n}\Big)^{nm}\prod_{i=0}^{n-1}\prod_{j=0}^{m-1}\frac{\gamma(-(j+1)nb^2+iv)}{\gamma(ba_1+jnb^2+iv)\gamma(ba_2+jnb^2+iv)\gamma(ba_3+jnb^2+iv)}\nonumber
\end{eqnarray}
\ \ \ \ \ \ \ \ \ \ \ \ \ \ \ \  \ \ \ \ \ \ \ \ \ \ \ \ \ \ \ \ \ \ \ \ \ \ \ \ \  \ \ \ \ $ba_1+ba_2+ba_3=v-nmb^2$;\\
where $x_1=0$, $x_2=(1,0)$, $x_3=\infty$ and the expectation value in (\ref{99}) is taken over the state that is the direct product of the ${\mathbb{Z}}_n$ parafermionic theory vacuum and the Fock vacuum for the massless field $\varphi$.

\end{document}